\documentclass[prl,aps,twocolumn,floatfix,tightenlines,amsmath,amssymb,showpacs,superscriptaddress,notitlepage,longbibliography]{revtex4-2}

\usepackage{bm}
\usepackage[pdftex]{graphicx,hyperref}
\hypersetup{colorlinks = true, urlcolor = blue, linkcolor = blue, citecolor = blue}
\usepackage{mathtools}
\usepackage{ulem}

\begin{document}

\title{Breakdown of Stoner Ferromagnetism by Intrinsic Altermagnetism}

\author{Chen Lu}
\email{luchen@hznu.edu.cn}
\affiliation{School of Physics and Hangzhou Key Laboratory of Quantum Matter, Hangzhou Normal University, Hangzhou 311121, China}
\author{Chao Cao}
\affiliation{Center for Correlated Matter and School of Physics, Zhejiang University, Hangzhou 310058, China}
\author{Huiqiu Yuan}
\affiliation{Center for Correlated Matter and School of Physics, Zhejiang University, Hangzhou 310058, China}
\author{Piers Coleman}
\affiliation{Center for Materials Theory, Department of Physics and Astronomy, Rutgers University, 136 Frelinghuysen Rd., Piscataway, NJ 08854-8019, USA}
\author{Lun-Hui Hu}
\email{lunhui@zju.edu.cn}
\affiliation{Center for Correlated Matter and School of Physics, Zhejiang University, Hangzhou 310058, China}

\begin{abstract}
The Stoner criterion for ferromagnetism arises from interaction-driven asymmetric filling of spin bands, requiring that the spin susceptibility: (i) peaks dominantly at $\mathbf{Q}=\bm{0}$; and (ii) diverges at a critical interaction strength. Here, we demonstrate that this Stoner mechanism breaks down due to competition with altermagnetic orders, even when both conditions are met. Altermagnetism in solids is characterized by collinear antiparallel spin alignment that preserves translational symmetry, and inherently fulfills these requirements. As a proof of concept, we study a two-orbital Hubbard model with electron filling near Van Hove singularities at high-symmetry momenta. Our results reveal that orbital-resolved spin fluctuations, amplified by strong inter-orbital hopping, stabilize intrinsic altermagnetic order. A quantum phase transition from altermagnetism to ferromagnetism occurs at critical Hund's coupling $J_H$. We further propose directional spin conductivity anisotropy as a detectable signature of this transition via non-local spin transport. This work establishes the pivotal role of altermagnetism in correlated systems.
\end{abstract}

\maketitle

\textit{Introduction--}
Altermagnetism (AM) represents a recently identified collinear magnetic phase that fundamentally challenges conventional classifications of quantum materials~\cite{naka2019spin,ahn2019antiferromagnetism,hayami2019momentum,vsmejkal2020crystal,yuan2020giant,shao2021spin,mazin2021prediction,ma2021multifunctional,yuan2021prm,vsmejkal2022beyond}. In real space, AM features compensated spin sublattices with antiparallel spins coupled via crystalline symmetries~\cite{vsmejkal2022emerging,bai2024altermagnetism,jungwirth2024altermagnets,fender2025altermagnetism}. Symmetry breaking under composite operations, such as parity-time reversal or translation-time reversal, generates momentum-dependent spin-splitting bands with distinctive $d$-, $g$-, or $i$-wave textures among others~\cite{vsmejkal2022emerging,bai2024altermagnetism,jungwirth2024altermagnets,fender2025altermagnetism,Fernandes2024prb}. Both $d$ and $g$ wave AMs have been experimentally confirmed in multiple quantum materials~\cite{feng2022anomalous,fedchenko2024ruo2,lin2024ruo2,gonzalez2023MnTe,krempasky2024MnTe,lee2024MnTe,osumi2024MnTe,liu2024chiral,reimers2024CrSb,ding2024CrSb,yang2025three,zhang2025crystal,jiang2024discovery}. The non-relativistic spin-splitting was also predicted through spin-channel Pomeranchuk instabilities~\cite{Hirsch1990prb,wu2004dynamic,wu2007fermi} and $d$-wave spin-density wave~\cite{Ikeda1998prl}. These spin-split metallic states enable novel spintronic functionalities despite vanishing net magnetization~\cite{wu2007fermi,vsmejkal2022emerging,gonzalez2021efficient,vsmejkal2022giant,guo2024emerging,shao2024antiferromagnetic,song2025altermagnets,Ouassou2023prl,zhang2024finite}.

AM is a distinct magnetic order beyond conventional ferromagnetic (FM) and N\'eel antiferromagnetic (AFM) paradigms. Recent advances reveal multiple pathways to AM: (i) spontaneous symmetry breaking in correlated electronic systems~\cite{maier2023weak,leeb2024spontaneous,Roig2024prb,das2024realizing,Sato2024prl,yu2025altermagnetism,Zhao2025prb,wang2024submit}, (ii) stacking-engineered van der Waals ferromagnetic heterostructures~\cite{He2023prl,liu2024twisted}, and (iii) AFM phase in non-centrosymmetric systems like ferroelectrics~\cite{vsmejkal2024altermagnetic,duan2025antiferroelectric,gu2025ferroelectric}. Extrinsic control via strain engineering, lattice vacancies, or spin clusters further expands the AM phase space~\cite{chakraborty2024strain,zhu2025design,li2025pressure}. Nevertheless, a fundamental question persists: what role does AM play in reshaping established theories of magnetic ordering?

\begin{figure}[t]
\centering
\includegraphics[width=0.85\linewidth]{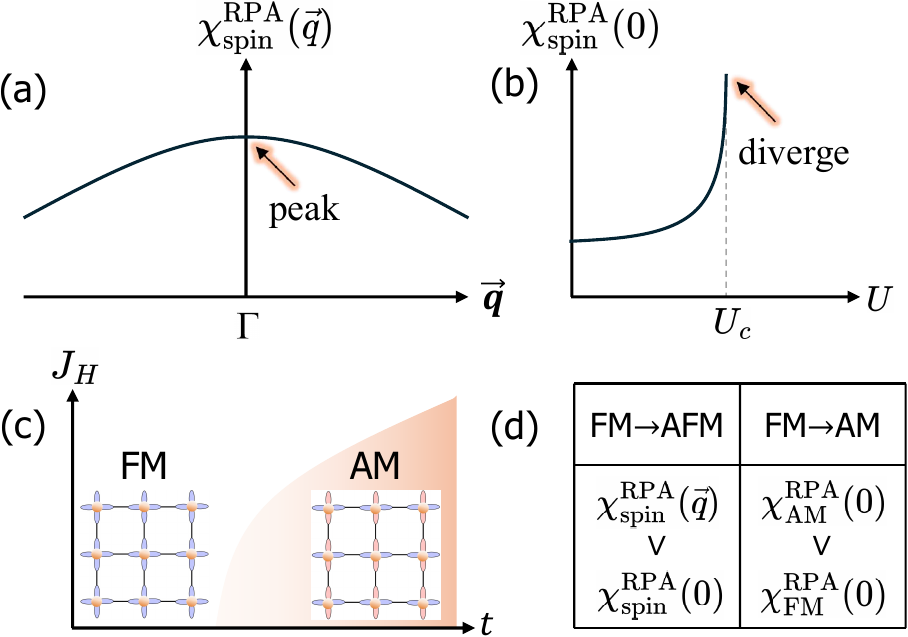}
\caption{Breakdown of Stoner FM. 
(a) First Stoner condition: dominant spin susceptibility at $\bm{q}=\bm{0}$. 
(b) Second Stoner condition: susceptibility divergence at critical $U_c$.
(c) Illustration for the competition between FM and AM in the $t$-$J_H$ phase diagram.
(d) Dual breakdown mechanisms of Stoner FM: (i) FM $\to$ N\'eel AFM transition when $\chi_{\text{spin}}^{\text{RPA}}(\bm{q}) > \chi_{\text{spin}}^{\text{RPA}}(\bm{0})$ (established mechanism~\cite{moriya2012spin,coleman2015introduction}); and (ii) FM $\to$ AM transition when $\chi_{\text{AM}}^{\text{RPA}}(\bm{0}) > \chi_{\text{FM}}^{\text{RPA}}(\bm{0})$ (this work).
A complete theory of Stoner FM must account for both breakdown channels.
}
\label{fig0}
\end{figure}

In this work, we establish intrinsic AM as a new mechanism that breaks down the Stoner paradigm for FM. The Stoner criterion requires~\cite{moriya2012spin,coleman2015introduction}: (i) dominant spin susceptibility at $\bm{q}=\bm{0}$ [Fig.~\ref{fig0}(a)] and (ii) susceptibility divergence at critical $U_c$ [Fig.~\ref{fig0}(b)]. While N\'eel AFM intrinsically violates this paradigm through susceptibility divergence at $\bm{q} \neq \bm{0}$, AM satisfies both conditions yet preempts FM ordering. Consequently, conditions (i) and (ii) are insufficient to guarantee FM when accounting for AM-FM competition [Figs.~\ref{fig0}(c-d)]. To resolve this paradox, we introduce criterion (iii): the relative divergence rates of $\chi_{\text{AM}}^{\text{RPA}}(\bm{0})$ and $\chi_{\text{FM}}^{\text{RPA}}(\bm{0})$ near $U_c$. As a proof of concept, we consider a two-orbital Hubbard model at electron fillings near Van Hove singularities (VHS) at high-symmetry momenta. We find that AM emerges spontaneously before FM due to inter-orbital hopping-amplified fluctuations. By tuning either interaction or hopping parameters, we find a quantum phase transition from FM to AM. We further propose directional spin conductivity anisotropy as a detectable signature of this transition, i.e., the breakdown of Stoner FM.

\textit{Minimal model with VHS--} 
To systematically identify intrinsic AM in competition with other magnetic phases including N\'eel AFM and FM, we study a minimal two-orbital Hubbard model on a square lattice. The system consists of two degenerate orbitals (e.g.,~$d_{xz}$ and $d_{yz}$) with the tight-binding Hamiltonian
\begin{align}\label{Hamiltonian_tb}
{\cal H}_{0}(\bm{k}) &= \varepsilon_0(\bm{k}) \tau_0 + \varepsilon_1(\bm{k}) \tau_x + \varepsilon_2(\bm{k}) \tau_z,
\end{align}
where $\varepsilon_0(\bm{k}) = -2t_0[\cos(k_x)+\cos(k_y)] - 2t_1[\cos(2k_x)+\cos(2k_y)]-\mu$ with $\mu$ the chemical potential, $\varepsilon_1(\bm{k})=4t_2\sin(k_x)\sin(k_y)$, $\varepsilon_2(\bm{k})=2t_0'[\cos(k_x) - \cos(k_y)] + 2t_1'[\cos(2k_x)-\cos(2k_y)]$, and $\tau_{x,y,z}$ are Pauli matrices acting on the orbital degrees of freedom. Our model Hamiltonian is expressed in the unit-cell gauge.  Among hopping parameters, $\{t_0,t_1\}$ are orbital-independent, while $\{t_0',t_1',t_2\}$ are orbital-dependent. The two bands are $\varepsilon_{\bm{k}}^{\pm} = \varepsilon_0(\bm{k}) \pm \sqrt{[\varepsilon_1(\bm{k})]^2 + [\varepsilon_2(\bm{k})]^2}$. In the limit $t_0'=t_1'=t_2=0$, Eq.~\eqref{Hamiltonian_tb} reduces to a single-band model where VHSs occur at the high-symmetry points $\mathbf{X}$ and $\mathbf{Y}$, since $\varepsilon_0(\mathbf{X}/\mathbf{Y}+\bm{k}) \propto \mp(t_0+4t_1)(k_x^2 - k_y^2)$. At half-filling with $t_1=0$, perfect particle-hole channel nesting emerges at $\mathbf{Q}=(\pi,\pi)$. This nesting is known to favor N\'eel AFM as a leading instability instead of FM, demonstrating the breakdown of Stoner criterion~\cite{moriya2012spin,coleman2015introduction}. Increasing $t_1$ reshapes the Fermi surface and thus disrupts perfect nesting, thereby suppressing AFM and stabilizing FM [see Sec.~A in Supplementary Material (SM)~\cite{sm2025}]. This underscores the critical role of VHSs near the Fermi energy on electronic properties.

\begin{figure}[t]
\centering
\includegraphics[width=0.48\textwidth]{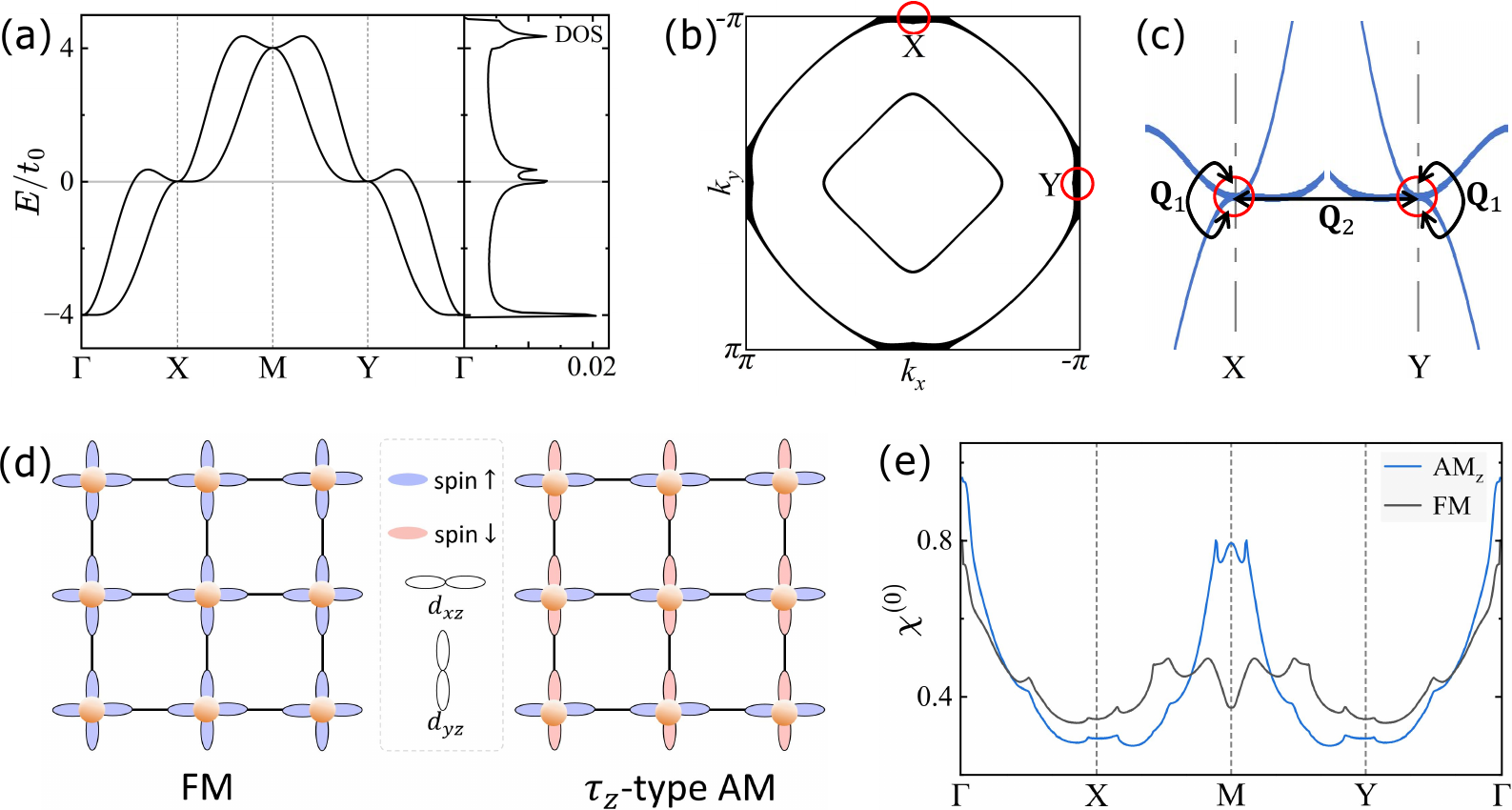}
\caption{Fermi surface, band structure and $\chi ^{(0)}(\bm{k})$ in the two-orbital square lattice model. (a) Band structure and the corresponding density of states (DOS) distribution. (b) Fermi surface in the first Brillouin zone, showing density-of-states ``hot spots'' near high-symmetry points $\textbf{X}$ and $\textbf{Y}$ due to VHS. (c) Corresponding schematic band structure near the Fermi level. The nesting vector $\bm{Q}_1=(0,0)$, $\bm{Q}_2=(\pi,\pi)$. (d) Real-space configurations of ferromagnetic order and $\tau_z$-type altermagnetic order.
(e) Bare susceptibilities in the $\tau_z$-type altermagnetic and ferromagnetic channels along the high-symmetry paths in the first Brillouin zone.}
\label{Lattice}
\end{figure}

We next turn on inter-orbital hoppings using the parameter set (with $t_0=1$ as energy unit): $t_1=0.09$, $t_2=0.77$, $t_0'=0$, $t_1'=0.36$, and chemical potential $\mu=-0.37$. These terms open a band gap while preserving twofold degeneracy at $\mathbf{X}$ and $\mathbf{Y}$ [Fig.~\ref{Lattice}(a)]. The Fermi surfaces in Fig.~\ref{Lattice}(b) reveal persistent VHSs at these points, where density of states accumulation forms pronounced ``hot spot'' [black dots]. These hot spots are connected by two dominant nesting vectors: $\mathbf{Q}_1 = (0, 0)$ (intra-VHS) and $\mathbf{Q}_2 = (\pi, \pi)$ (inter-VHS), as illustrated in Fig.~\ref{Lattice}(c). While $\mathbf{Q}_1$ supports Stoner FM, $\mathbf{Q}_2$ corresponds to N\'eel AFM. Crucially, $\mathbf{Q}_1$ also permits AM, characterized by antiparallel spins locked to distinct atomic orbitals. Real-space configurations of these magnetic orders appear in Fig.~\ref{Lattice}(d). To resolve competition among these orders, we calculate the bare susceptibility tensor in the momentum-frequency space,
\begin{align}\label{chi0}
\begin{split}
&[\chi^{(0)}(\bm{k},i \omega)]^{l_1l_2}_{l_3l_4} \equiv \frac{1}{N} \sum_{\bm{k}_1\alpha \beta}  [\xi^{\alpha}_{l1}(\bm{k}_1)]^\ast \xi^{\beta}_{l2}(\bm{k}_1+\bm{k}) \times \\
&\qquad\quad [\xi^{\beta}_{l3}(\bm{k}_1+\bm{k})]^\ast  \xi^{\alpha}_{l4}(\bm{k}_1)\frac{\eta_F(\varepsilon^{\beta}_{\bm{k}_1+\bm{k}})-\eta_F(\varepsilon^{\alpha}_{\bm{k}_1})}{i\omega+\varepsilon^{\alpha}_{\bm{k}_1}- \varepsilon^{\beta}_{\bm{k}_1+\bm{k}}},
\end{split}
\end{align}
where $l_1, l_2, l_3, l_4$ are orbital indices, $\alpha, \beta$ are band indices, $N$ is the lattice size, $\varepsilon^{\alpha}_{\bm{k}}$ and $\xi^{\alpha}(\bm{k})$ are the $\alpha$-th eigenvalue and eigenvector of ${\cal H}_{0}(\bm{k})$, respectively, and $\eta _F$ is the Fermi-Dirac distribution function. With spin-orbit coupling excluded, Eq.~\eqref{chi0} contains no explicit spin index. We then define static bare susceptibilities for both FM and AM channels as~\cite{Roig2024prb},
\begin{align}\label{chiO}
\chi^{(0)}_{\alpha}(\bm{k}) = \frac{1}{2} \sum_{l_1l_2l_3l_4}  [\bar{\cal O}_\alpha]_{l_1l_2} [\bar{\cal O}_\alpha]_{l_3l_4} [\chi^{(0)}(\bm{k},0)]^{l_1l_2}_{l_3l_4},
\end{align}
with $\bar{\cal O}_{\text{FM}} = \tau_0$ and $\bar{\cal O}_{\text{AM}} = \tau_z$. The case with $\bar{\cal O}_{\text{AM}}=\tau_x$ could be formally mapped to the $\tau_z$-type AM via a unitary transformation and can be realized with an alternative set of parameters [see Sec.~B in SM~\cite{sm2025}]. The calculated $\chi^{(0)}_{\text{FM}/\text{AM}}(\bm{k})$ along high-symmetry lines are shown in Fig.~\ref{Lattice}(e), revealing prominent peaks around both nesting vectors: $\mathbf{Q}_1$ and $\mathbf{Q}_2$. As expected, finite $t_1$ leads to dominant FM rather than N\'eel AFM, with $\chi^{(0)}_{\text{FM}}(\Gamma)>\chi^{(0)}_{\text{FM}}(\mathbf{M})$ [black curve]. Remarkably, the AM susceptibility at $\Gamma$ exceeds the FM value, with $\chi^{(0)}_{\text{AM}}(\Gamma)>\chi^{(0)}_{\text{FM}}(\Gamma)$ [blue curve]. This demonstrates AM-induced breakdown of Stoner FM, an effect that remains robust against electronic interactions as shown later.

\begin{figure*}[t]
\centering
\includegraphics[width=0.88\textwidth]{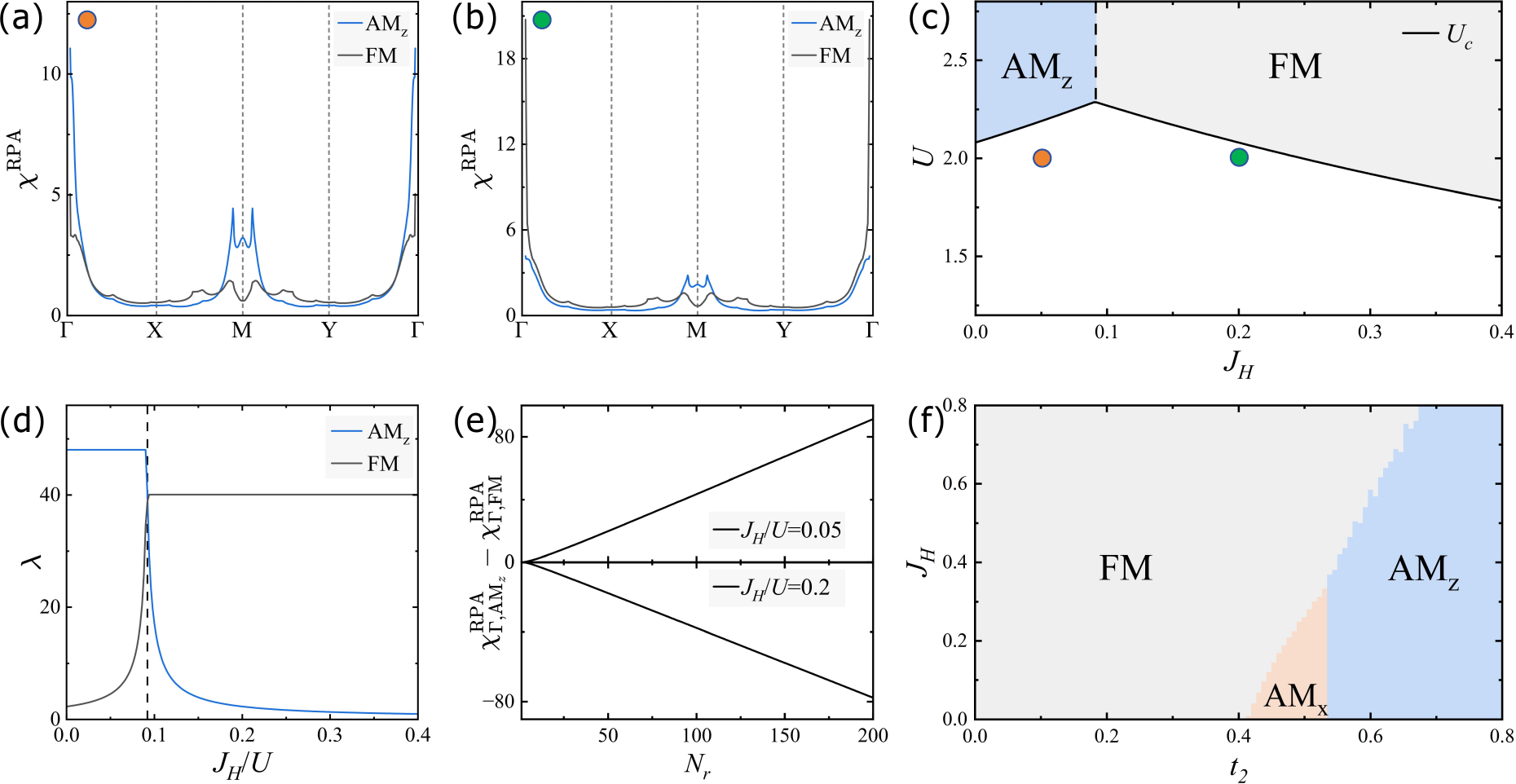}
\caption{Comparison of susceptibilities between different magnetic orders and the corresponding interaction phase diagram at the RPA level. (a) High-symmetry-line plots of $\tau_z$-type AM ($\text{AM}_z$) and FM susceptibilities with $J_H/U=0.05$ and $U=2$. (b) Same as (a) for $J_H/U=0.2$. (c) Phase diagram showing three distinct regimes: the altermagnetic phase ($\text{AM}_z$,blue), ferromagnetic phase (FM, gray), and non-magnetic phase (white). The magnetic-to-non-magnetic boundary is marked by the critical $U_c$ (solid black curve), while the dashed line ($J_H/U=0.092$) separates the $\tau_z$-type AM and FM phases.
(d) The two dominant eigenvalues of the susceptibility matrix $[\chi^{\text{RPA}}_\text{spin}(0)]^{l_1l_1}_{l_2l_2}$ correspond to two magnetic phases: $\text{AM}_z$ and FM. Their eigenvalues $\lambda$ evolve with the ratio $J_H/U$, tracking the transition at $J_H/U=0.092$ between these two phases.
(e) The divergent behavior of ${\chi}^{\text{RPA}}_{\Gamma,\text{AM}_z}-{\chi}^{\text{RPA}}_{\Gamma,\text{FM}}$ as $U$ approaches $U_c$, where $U=(1-1/N_r)U_c$ with $N_r$ increasing from $0$ to $200$.
Results are shown for $J_H/U=0.05$ and $J_H/U=0.2$. 
All calculations use the same parameters as in Fig.~\ref{Lattice}(a). (f) Phase diagram in the $t_2$–$J_H$ plane calculated at $U = 2$ with band parameters $\{t_0, t_1, t^{\prime}_1, t_2, \mu\} = \{1, 0.15, 0.6, 0.77, -0.625\}$.
}
\label{O4}
\end{figure*}

This breakdown requires maximal susceptibility divergence at $\mathbf{Q}=\bm{0}$, a condition whose critical role was unrecognized prior to the discovery of AM. Its direct application identifies $336$ altermagnetic candidates among $2199$ magnetic materials~\cite{wan2024high}. We now examine the role of VHSs on the $\mathbf{Q}=\bm{0}$ divergence. As specified by Eq.~\eqref{chi0}, the Lindhard function term $[\eta_F(\varepsilon^{\beta}_{\bm{k}_1+\mathbf{Q}})-\eta_F(\varepsilon^{\alpha}_{\bm{k}_1})]/[\varepsilon^{\alpha}_{\bm{k}_1}- \varepsilon^{\beta}_{\bm{k}_1+\mathbf{Q}}]$ can diverge while the eigenvector product remains finite. At low temperatures, divergence at $\mathbf{Q} = \bm{0}$ requires: (1) a finite numerator, requiring that the $\alpha$ and $\beta$ bands to be on opposite sides of the Fermi level (one occupied, one unoccupied); (2) a vanishing denominator, demanding both band energies approach the Fermi level; and (3) the sufficiently high density of states around $\bm{k}_1$-points where conditions (1) and (2) are concurrently met, ensuring the integrated susceptibility yields a strong divergence. Moreover, the condition (3) is automatically fulfilled when the Fermi level lies near high-symmetry point VHSs, as illustrated in Fig.~\ref{Lattice}(c).

\textit{Phase diagrams--}
Although Fig.~\ref{Lattice}(e) indicates that intrinsic AM disrupts Stoner FM via the susceptibility relationship $\chi^{(0)}_{\text{AM}}(\Gamma) > \chi^{(0)}_{\text{FM}}(\Gamma)$, this breakdown mechanism requires further scrutiny when electronic interactions are present. To investigate this, we map the phase diagram of competing FM and AM orders governed by the repulsive Hubbard-Hund Hamiltonian,
\begin{align} \label{Hamiltonian}
\begin{split}
H_{int}&=U\sum _{\bm{i},\tau} n_{\bm{i}\tau \uparrow }n_{\bm{i}\tau \downarrow }+V\sum_{\bm{i}s,s^{\prime}} n_{\bm{i},x,s}n_{\bm{i},y,s^{\prime}} \\
&+J_H \sum _{\bm{i}}  \sum _{s,s^{\prime}} c^{\dagger}_{\bm{i},x,s}c^{\dagger}_{\bm{i},y,s^{\prime}}c_{\bm{i},x,s^{\prime}}c_{\bm{i},y,s}  \\
&+J_H \sum _{\bm{i}} c^{\dagger}_{\bm{i},x,\uparrow} c^{\dagger}_{\bm{i},x,\downarrow} c_{\bm{i},y,\downarrow} c_{\bm{i},y,\uparrow} + h.c.,
\end{split}
\end{align}
where $c_{\bm{i},\tau,s}$ is the electron annihilation operator at site $\bm{i}$ with orbital $\tau$ and spin $s$, $n_{i\tau s}=c_{\bm{i},\tau,s}^\dagger c_{\bm{i},\tau,s}$ is the density operator, $\tau=\{x,y\}$ labels the $\{d_{xz}, d_{yz}\}$ orbitals, and $s=\{\uparrow,\downarrow\}$ denotes the spin. Here, $U$ is the intra-orbital Hubbard interaction, $V$ is the inter-orbital Hubbard term, and $J_H$ is the Hund's coupling. The spin rotation symmetry imposes the constraint $U=V+2J_H$ \cite{castellani1978magnetic}. We then solve ${\cal H}_{0}(\bm{k})+H_{int}$ to explore the phase diagrams, by employing the standard multi-orbital random-phase approximation (RPA) approach~\cite{scalapino1986d,hamann1969properties,Roig2024prb}. Within the RPA framework, repulsive onsite Hubbard interactions suppress charge susceptibility while enhancing spin susceptibility~\cite{scalapino1986d}. The RPA-renormalized static susceptibility for FM or AM orders is given by
\begin{align} \label{chiO_RPA}
\chi_{\alpha}^{\text{RPA}}(\bm{k}) &= \frac{1}{2} \sum_{l_1l_2l_3l_4}  [ \bar{\cal O}_\alpha ]_{l_1l_2} [  \bar{\cal O}_\alpha ]_{l_3l_4} [\chi^{\text{RPA}}_{\text{spin}}(\bm{k})]^{l_1l_2}_{l_3l_4} ,  
\end{align}
where $[\chi^{\text{RPA}}_{\text{spin}}(\bm{k})]^{l_1l_2}_{l_3l_4}$ denotes the static spin susceptibility tensor and is determined by the Dyson equation, $\chi^{\text{RPA}}_{\text{spin}}(\bm{k}) = [I - \chi^{(0)}(\bm{k}){\cal U}_{s}]^{-1} \chi^{(0)}(\bm{k})$, where $I$ denotes the identity matrix and ${\cal U}_{s}$ represents the spin-channel interaction matrix [see Sec.~C in SM~\cite{sm2025}]. The divergence of $\chi^{\text{RPA}}_{\text{spin}}(\bm{k})$ at critical interaction strength $U_c$ and ordering wavevector $\bm{Q}$ generally signals an instability: $U_c$ establishes the magnetic phase boundary, while $\bm{Q}$ determines the real-space periodicity of the dominant magnetic order. Crucially, Eq.~\eqref{chiO_RPA} indicates that the divergence of $\chi^{\text{RPA}}_{\text{spin}}(\bm{k})$ at $U_c$ simultaneously triggers the divergence of $\chi_{\text{FM/AM}}^{\text{RPA}}(\bm{k})$. For $U > U_c$, long-range magnetic order develops with $\bm{Q} = \mathbf{0}$ corresponding to uniform FM or AM configurations. Figure~\ref{O4} presents results based on Eq.~\eqref{chiO_RPA}: panel (a) shows $\chi_{\text{AM}}^{\text{RPA}}(\Gamma) > \chi_{\text{FM}}^{\text{RPA}}(\Gamma)$ at small $J_H$ ($U=2$, $J_H/U=0.05$), while panel (b) demonstrates the reversed relationship at large $J_H$ ($U=2$, $J_H/U=0.2$). It indicates that while the bare susceptibility is maximized in the AM channel, the RPA-enhanced susceptibility only exceeds the FM channel in the small $J_H$ regime.

We then map out the complete phase diagram in the $U$–$J_H/U$ plane [Fig.~\ref{O4}(c)]. The solid black curve marks the boundary between non-magnetic (white) and magnetic states (blue and gray), tracing $U_c$ as a function of $J_H/U$. Within the magnetic regime, we identify two distinct phases: the $\tau_z$-type AM state (blue) and the FM phase (gray). A direct quantum phase transition separates these orders at $J_H/U \approx 0.1$, indicated by the dashed black curve. This critical point is established via two complementary approaches: (i) analyzing the eigenvectors of the spin susceptibility matrix $[\chi^{\text{RPA}}_{\text{spin}}(\Gamma)]^{l_1l_1}_{l_2l_2}$ [Fig.~\ref{O4}(d)], and (ii) comparing the divergence rate of $\chi_{\text{AM}/\text{FM}}^{\text{RPA}}(\Gamma)$ [Fig.~\ref{O4}(e)]. We first diagonalize this matrix and extract the two dominant eigenvalues $\lambda_{\text{AM}}$ and $\lambda_{\text{FM}}$, which correspond to the $\tau_z$-type AM and FM order parameters, respectively. At $U=0.99U_c$, $\lambda_{\text{AM}}$ and $\lambda_{\text{FM}}$ exhibit a clear crossing at $J_H/U=0.092$, signaling the transition. As anticipated, the AM phase is stabilized only at weak Hund's coupling ($J_H$). This behavior is further corroborated near criticality through the susceptibility difference $\Delta\chi = \chi_{\text{AM}}^{\text{RPA}}(\Gamma) - \chi_{\text{FM}}^{\text{RPA}}(\Gamma)$. By using a scaling approach $U=(1-1/N_r)U_c$ with $N_r: 0\to 200$, we obtain $\Delta\chi \geq 0$ in the weak $J_H/U$ regime. These results demonstrate the robustness of the AM-induced breakdown of Stoner FM at small $J_H$.

The competition between Stoner FM and intrinsic AM arises from the energy balance between the kinetic contribution (inter-orbital hopping) and Hund's coupling $J_H$, where $t_2$ in Eq.~\eqref{Hamiltonian_tb} could mediate an effective inter-orbital AFM-type Heisenberg coupling~\cite{wang2024submit}. The parameters set $\{t_0',t_1'\}$ maintain similar roles as $t_2$ under the $\tau_x \leftrightarrow \tau_z$ transformation. Using Eq.~\eqref{chiO_RPA}, we identify the most strongly RPA-enhanced susceptibility among the FM, $\tau_z$-type AM, $\tau_x$-type AM channels. As expected, $\tau_x$-type AM dominates as the leading instability at small $J_H$ and $t_2$. Increasing $t_2$ shifts the dominant instability to $\tau_z$-type AM, while enhanced $J_H$ consistently favors Stoner FM in both regimes. This competition is mapped in the $t_2$-$J_H$ phase diagram shown in Fig.~\ref{O4}(f).

\begin{figure}[t]
\centering
\includegraphics[width=0.48\textwidth]{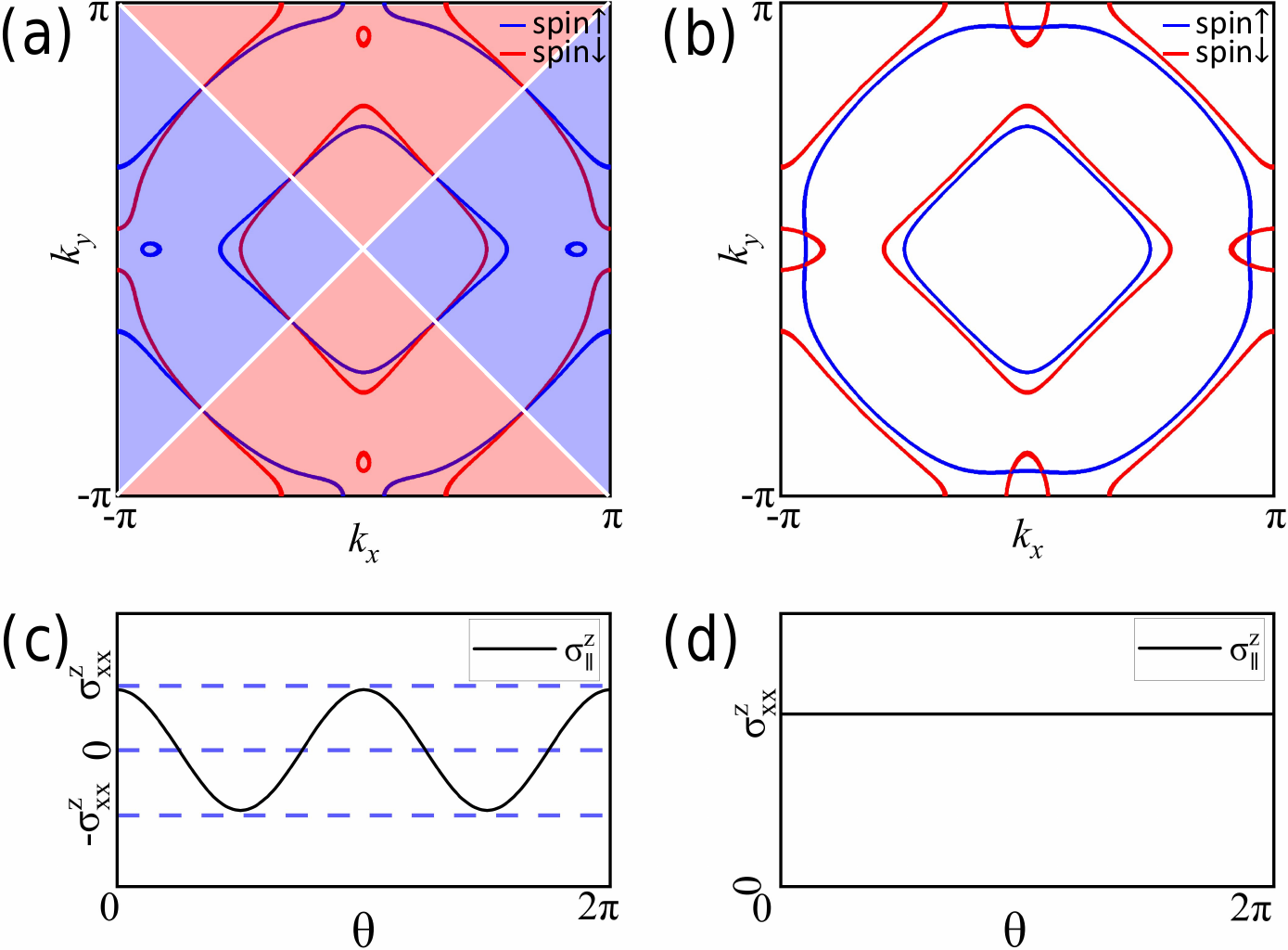}
\caption{Spin conductivity anisotropy as a probe of quantum phase transition.
(a) Fermi surface with $d$-wave spin splitting induced by the $\tau_z$-type AM order with $\Delta_{\text{AM}}=0.2$.
(b) Fermi surface with isotropic spin splitting from the ferromagnetic order with $\Delta_{\text{FM}}=0.2$.
(c) Anisotropic spin conductivity $\sigma_{\parallel}(\theta)$ for $\tau_z$-type AM, following $\sigma_{\parallel}(\theta)=\sigma^z_{xx}\cos(2\theta)$ with $\sigma^z_{xx}=-\sigma^z_{yy}=8.83$, characteristic of $d$-wave symmetry.
(d) Nearly isotropic spin conductivity for FM, demonstrating distinct transport signatures between phases.
All calculations use the same parameters as in Fig.~\ref{Lattice}(b).
}
\label{fig4}
\end{figure}

\textit{Experimental signatures--}
We now discuss experimentally verifiable signatures of the quantum phase transition between Stoner FM and intrinsic AM. The mean-field Hamiltonian is given by,
\begin{align}\label{Hamiltonian_MF}
\begin{split}
{\cal H}_{\text{MF}}(\bm{k})&= {\cal H}_{0}(\bm{k})s_0 + \Delta_{\text{AM}}\tau_zs_z  + \Delta_{\text{FM}}\tau_0s_z,
\end{split}
\end{align}
where $\Delta_{\text{AM}}$ and $\Delta_{\text{FM}}$ are the respective order parameters. The AM-driven breakdown of Stoner FM corresponds to the transition from $(\Delta_{\text{AM}}=0, \Delta_{\text{FM}}\neq 0)$ to $(\Delta_{\text{AM}}\neq 0, \Delta_{\text{FM}}= 0)$. The distinct broken symmetries suggest this transition is likely first-order, manifesting through two key signatures. (i) Band structure anisotropy: the AM phase exhibits $d$-wave band splitting with spin-momentum locking [Fig.~\ref{fig4}(a)], contrasting sharply with the isotropic spin splitting and uniform spin polarization of the FM phase [Fig.~\ref{fig4}(b)]. (ii) The AM phase generates $\sin(2\theta)$-modulated spin conductivity from its $d$-wave spin texture, while the FM phase displays crystal-orientation-independent isotropic spin transport.

To quantify the spin current response, we compute the spin conductivity within the Kubo-Streda formalism~\cite{freimuth2014spin},
\begin{align} \label{sigma}
\sigma^{a}_{bc} = \text{Re} \sum_{\bm{k},\alpha,\beta} \frac{\langle \xi^{\beta}(\bm{k}) | \hat{J}^a_b | \xi_{\alpha}(\bm{k}) \rangle \langle \xi_{\alpha}(\bm{k}) | \hat{v}_c | \xi^{\beta}(\bm{k}) \rangle }{[(\mu- \varepsilon^{\alpha}_{\bm{k}})^2+\eta^2][(\mu-\varepsilon^{\beta}_{\bm{k}})^2+\eta^2]},
\end{align}
in unit of $e \hbar \eta^2 / (N \pi)$. Here $\hat{v}$ is the velocity operator, and the spin-current operator $\hat{J}^a_b=\frac{1}{2}\left \{ \hat{s}_a,\hat{v}_b \right \}$. We adopt natural units ($e = \hbar = 1$) and set the quasiparticle broadening $\eta = 0.02$. We focus on the longitudinal spin conductivity $\sigma^z_{\parallel}$ driven by a static electric field $\vec{E} \parallel (\cos\theta, \sin\theta)$, where $\theta$ defines the field orientation. Figure~\ref{fig4}(c) reveals a hallmark signature of the $d$-wave AM phase: $\sigma^z_{\parallel}(\theta) \propto \cos(2\theta)$ modulation with sign reversals at $\theta = n\pi/2$ ($n\in\mathbb{Z}$). This four-fold symmetric response per $2\pi$ rotation directly originates from the $d_{xy}$-symmetric spin-split Fermi surface [Fig.~\ref{fig4}(a)]. This behavior stands in sharp contrast to the ferromagnetic phase [Fig.~\ref{fig4}(d)], where $\sigma^z_{\parallel}$ remains nearly constant with $2\pi$ periodicity.

\textit{Conclusion--}
In summary, we have demonstrated that the conventional Stoner paradigm for ferromagnetism fundamentally breaks down due to competition with altermagnetic orders. While the Stoner criterion requires (i) dominant spin susceptibility at $\mathbf{Q} = \mathbf{0}$ and (ii) divergence at a critical interaction strength $U_c$, we show these conditions are insufficient to guarantee ferromagnetic ordering when altermagnetism is present. Through a minimal two-orbital Hubbard model near Van Hove singularities, we establish that orbital-resolved spin fluctuations can stabilize intrinsic altermagnetic order via strong inter-orbital hopping. Crucially, we identify a third criterion governing this breakdown: the relative divergence rates of $\chi^{\text{RPA}}_{\text{AM}}(\mathbf{0})$ and $\chi^{\text{RPA}}_{\text{FM}}(\mathbf{0})$ near $U_c$. When $\chi^{\text{RPA}}_{\text{AM}}$ diverges faster than $\chi^{\text{RPA}}_{\text{FM}}$, altermagnetism preempts ferromagnetism despite both orders sharing the $\mathbf{Q}=\mathbf{0}$ instability channel. This mechanism is distinct from the established breakdown via N\'eel antiferromagnetism (divergence at $\mathbf{Q} \neq \mathbf{0}$) and highlights altermagnetism as a generic competitor to Stoner ferromagnetism in correlated multi-orbital systems.
Our main conclusion can be generalized to systems where atomic orbitals are replaced by non-equivalent sublattices within a unit cell.

The orbital-active magnetic materials may provide a platform to verify our results, such as transition metal oxides where orbital order coexists with antiferromagnetism, such as: $\alpha-$Sr$_2$CrO$_4$ with a 3$d^2$ electronic configuration in the Cr$^{4+}$ state~\cite{pandey2021origin,lee2022ultrafast}; SrRuO$_3$ thin films with SrO termination~\cite{autieri2016antiferromagnetic}; and various vanadium-based oxides V$_2$O$_3$~\cite{shiina2001atomic} and ZnV$_2$O$_4$~\cite{maitra2007orbital}. Moreover, the $\tau_z$-altermagnet exhibits a vanishing local magnetic moment while retaining a finite quadrupole order. This characteristic may be analogous to the phenomenology of hidden order in magnetic materials~\cite{Mydosh2011rmp}. Our work demonstrates that such elusive orders can be detected via their signature in spin-polarized bands and spin conductivity.

\vspace{\baselineskip}

\textit{Acknowledgments.--}
We thank A.~Gabriel, C.~X.~Liu, Y.~Z.~You, C.~Li, H.~K.~Jin, Z.~M.~Pan and F.~Yang for helpful discussions. We thank S.~B.~Zhang for careful reading of the manuscript. 
C.L. is supported by the National Natural Science Foundation of China under the Grants No. 12304180.
L.H.H. is supported by the start-up of Zhejiang University and the Fundamental Research Funds for the Central Universities (Grant No. 226-2024-00068).
This work has been supported by National Key R\&D Program of China (Grant No. 2022YFA1402200), the National Natural Science Foundation of China (Grants No. 12034017).

\bibliography{references}

\end{document}